\documentclass[aps,prd,showpacs,nofootinbib,preprint,tightenlines]{revtex4}

\def\be{\nopagebreak[3]\begin{equation}}
\def\ee{\end{equation}}
\def\ba{\nopagebreak[3]\begin{eqnarray}}
\def\ea{\end{eqnarray}}

\def\d{{\rm d}}

\def\half{{\textstyle{1\over2}}}

\newcommand{\teta}{\rlap{\lower2ex\hbox{$\,\tilde{}$}}\eta{}}

\begin{document}
\preprint{\vbox{\baselineskip=12pt \rightline{ICN-UNAM-05/01}
\rightline{gr-qc/0503078} }}
\title{Towards a new approach to Quantum Gravity Phenomenology}
\author{Alejandro Corichi}
\email{corichi@nucleares.unam.mx} \affiliation{Instituto de
Ciencias Nucleares,
 Universidad Nacional Aut\'onoma de M\'exico,\\
A. Postal 70-543, M\'exico D.F. 04510, M\'exico}
\author{Daniel Sudarsky}
\email{sudarsky@nucleares.unam.mx} \affiliation{Instituto de
Ciencias Nucleares,
Universidad Nacional Aut\'onoma de M\'exico,\\
A. Postal 70-543, M\'exico D.F. 04510, M\'exico}


\begin{abstract}
The idea that quantum gravity manifestations would be associated
with a violation of Lorentz invariance is very strongly bounded
and faces serious theoretical challenges. This leads us to
consider an alternative line of thought for such phenomenological
search. We discuss the underlying viewpoint and briefly mention
its possible connections with current theoretical ideas. We also
outline the challenges that the experimental search of the effects
would seem to entail.
\end{abstract}

\pacs{04.60.-m, 04.60.Pp, 04.80.-y 11.30.Cp.}
 \maketitle

\section{Introduction}

\noindent There has been recently a great deal of interest in
possible phenomenological manifestations of quantum gravity
effects \cite{Quant Fluct}. These effects are thought to arise
from the granularity of space-time at the Planck scale leading to
a breakdown  of Lorentz Invariance in its low energy limit and
would generally imply, in contrast with one of the most cherished
and useful principles of physics, the existence of a preferential
frame; a new version of the XIX${}^{th}$ century
Ether\footnote{Another possibility that has been considered in the
literature are deformations of the Lorentz Algebra or their
representations such as the so called Double Special Relativity
(DSR) proposals \cite{DSR}.  We will not deal with these ideas in
this manuscript.}. In fact a large collection of very tight bounds
have been obtained by considering astrophysical observations and
Laboratory experiments \cite{Bounds}. Moreover, a recent analysis
of the way that such a preferred frame granularity would affect
the radiative corrections in the standard model of particle
physics, shows that the basic suppositions are in conflict with
what we understand about quantum field theory together with even
low accuracy tests of Lorentz Invariance in particle physics
\cite{Us,radiative}.

The purpose of this paper is to explore what seems to be the next
most natural assumption regarding the way quantum gravity might
become manifest. We shall first motivate a proposal, starting with
a plausibility argument constructed to reconcile our intuitions
about quantum gravity, with the evidence against a Violation of
Lorentz Invariance (VLI) of the sort that has been considered. We
will then present the new possible ways in which an effective
theory might be the low energy limit of certain quantum gravity
theories, and finally we shall explore the possible
phenomenological implications of this new proposal.

Let us try to motivate the need for a new approach to the subject.
The idea is then that somehow the fact that the underlying
symmetry of the fundamental structure is itself the Lorentz
Symmetry, leads to a situation in which  the large scale Lorentz
Symmetry is protected by the symmetry of the fundamental granular
structure. In other words, that, given the symmetric nature of the
granular structure, the existence of a fundamental granularity
might not show up as an observational brake-down of the symmetry,
when the macroscopic physical entity (here the spacetime geometry)
is itself fully symmetric. Thus a region  of spacetime, normally
considered as well approximated by Minkowski metric, would not
manifests the granular structure of the quantum spacetime, through
the breakdown of Lorentz Invariance. This line of thought would
then explain naturally the previously mentioned empirical
evidence. The point, however, is not the explanation itself, which
admittedly is at best sketchy, but rather to motivate the next
lines of thought. In view of the above, we note that the only
interesting situation that would be left open to investigations is
that in which the macroscopic space-time that is to be probed is
not fully compatible with the symmetry of its basic constituents.
The idea is to think in analogy to  what happens when a large
crystal has the same symmetry (say cubic)  of the fundamental
crystal, one could expect no deviations from fully cubic symmetry,
as a result of the discrete nature of the fundamental building
blocks. However if one wants  to build a macroscopic crystal whose
global form is not compatible with the structure of the
fundamental crystals, say hexagonal,  the surface will necessarily
include some roughness, and thus a manifestation of the granular
structure, would occur through a breakdown of the  exact hexagonal
symmetry.

This simple picture will be guiding our analysis, and in that
respect we must keep in mind that we  will be referring to the
physical description of the situation at two levels. The first,
the effective or macroscopic level of description  would
correspond in the example of the physics of solids, to a
continuous description of physical objects, which we could
envision as being expressed in terms  of,  say, a smooth metric
and extrinsic curvatures of the surfaces of the solid,  and smooth
functions giving the mass density, pressures and viscosities in
its inside, etc. In the case of gravitation it would correspond to
general relativity. The second level would correspond to a
fundamental quantum mechanical solid state description of  both
the interior and surface constituents in the case of our solid
object, and a yet unknown theory of the underlying and somehow
granular structure of spacetime. The connection between the two
levels will be guided by symmetry arguments, and thus we will
refer often using the first level of description to the situations
in  which the granular nature of the underlying or fundamental
description would become manifest. We must keep this in mind for,
otherwise, statements involving simultaneously the two levels of
description would seem nonsensical.

Following with this line of thought and with the solid state
physics analogy, we would expect that it might be precisely in the
event of a  failure of the space-time to be exactly Minkowski in
an open domain, that the underlying granular structure of quantum
gravity origin, would become manifest, affecting the propagation
of  the various matter fields. The effective description of such
situation should thus involve the Riemann tensor, (which is known
to precisely describe the failure of a space-time to be Minkowski
over an open domain). This is, we know that in every point of
spacetime we can construct ``local  inertial frames" ,
corresponding to the Riemann Normal coordinates at that point, and
that the failure of such construction to provide an extended
inertial frame is measured by the Riemann tensor. Thus the
non-vanishing of Riemann would correspond to the macroscopic
description of the situation where the microscopic structure of
space-time might become manifest. Moreover, we can expect, due to
the implicit correspondence of the macroscopic description with
the more fundamental one, that the Riemann tensor  would also
indicate the space-time directions with which  the sought effects
would be associated. This would replace, in the current approach,
the global selection of a preferential reference frame that was
implicit on most schemes of --let us now refer to it as-- the old
approach towards Quantum Gravity Phenomenology (QGP).

In fact, when considering the phenomenology, it is important to
recall  that the Ricci tensor represents that part of the Riemann
tensor which, at least on shell, is locally determined by the
energy momentum of matter at the events of interest. Thus the
coupling of matter to the Ricci tensor part of the Riemann tensor
would, at the phenomenological level, reflect a sort of pointwise
self interaction of matter that would amount to a locally defined
renormalization of the usual phenomenological terms such as a mass
or kinetic term in the Lagrangian. However we are interested in
the underlying structure of space-time rather that the self
interaction of matter. Thus we would need to ignore the aspects
that are known to encode the latter, which in our case would
correspond to all lagrangian terms containing only that part of
the Riemann tensor proportional to the Ricci tensor,  coupled to
matter fields. The rest of the Riemann tensor can thus be thought
to reflect the aspect of local structure of spacetime associated
solely with the gravitational degrees of freedom. This local
structure, which is codified in the macroscopic theory by the Weyl
tensor, would in the microscopic theory reflect aspects of the
quantum gravitational structure of spacetime in itself. The
effects we are interested, which refer only to the gravitational
aspects, and  that would therefore constitute a probe into the
quantum mechanical nature of spacetime, would thus be associated
with the coupling of the Weyl tensor with the matter fields.In
everyday situations (i.e. in the absence of gravitational waves)
the Weyl tensor is also connected with the nearby ``matter
sources" but such connection involves the propagation of their
influence through the spacetime and thus the structure of the
latter would be playing a central role in the way the influences
become manifest. In this sense the Weyl tensor reflects the
``nonlocal effects" of the matter in contrast with the Ricci
tensor or curvature scalar that are determinable from the latter
in a completely local way.

In this paper we will start  the investigation of such ideas, in a
rather heuristic form, firstly because we do not want at this
point to commit ourselves to a specific proposal for the
quantization of gravitation, and secondly, because our first aim
is to outline the simplest  available options  for a more complex
realm of possibilities regarding quantum gravity phenomenology.
The main objective of this work is, therefore, to open the way to
a new perspective of analyzing (and possible observing) the
fundamental discreteness of the spacetime geometry.

The paper is organized as follows. In Section~\ref{sec:2} we will
study the most straight forward approach, showing that in this
scenario the situation regarding the observability of the quantum
gravity effects, is very pessimistic as the suppressions turn out
to be bigger than expected at first sight. In Section~\ref{sec:3}
we will explore a more complicated approach, that, on the one hand
looks as rather contrived, at least in the tensor  language, which
we must recognize,  might not have much to do with the natural
language that describes the quantum gravity realm, and on the
other hand seems to  yield relative large, and thus to a more
promising  scenario for the observability of  such effects. In
Section~\ref{sec:4} we shall give some arguments supporting our
general view, but geared towards its realization in a particular
approach to quantum gravity, the Loop Quantum Gravity Program. In
Section~\ref{sec:5} we will discuss the expected orders of
magnitude and other issues that would confront the experimental
investigation of these issues. We will end with a brief discussion
of our analysis and with some conclusions in Section~\ref{sec:6}.

\section{Straightforward Approach}
\label{sec:2}

\noindent The fundamental fields of the standard model are boson
and fermions, thus we start by  considering all couplings of such
fields to the Riemann tensor.  The Riemann tensor has mass
dimension 2, the fermions have mass dimension 3/2 and the bosons
have mass dimension 1. Naturally the dimension $n$ Lagrangian
terms would be suppressed by a factor $(1/M_{\rm
Planck})^{(n-4)}$.  Furthermore we will as usual assume observer
covariance and the absence of globally defined non-dynamical
tensor fields. A careful examination will reveal that the least
possible suppression corresponds to a unique term involving
coupling of fermions and the Riemann tensor.
\be
{\cal L}^\prime_{\rm f}=\frac{\xi }{M_{\rm Pl}}\, R_{\mu \nu
\rho \sigma}\,\bar \Psi\, \gamma^{\mu}  \gamma^{\nu} \gamma^{\rho
} \gamma^{\sigma}\, \Psi\, ,
\ee
where $\Psi$ stands for the various fundamental spinor fields in
the standard model, $\gamma^{\mu}$ for the Dirac matrices, $R_{\mu
\nu \rho \sigma}$ for the Riemann tensor, and $\xi$ a
dimensionless number characterizing the geometrical aspects of the
appropriate state of the quantum gravity theory and the fermions
interaction with it. Note in particular that one can not write a
similar dimension 5 operator involving boson fields. To proceed
further we note that we can write the product of the four gamma
matrices in terms of the symmetrized and  anti-symmetrized
products leading to an expansion containing only the matrices
$\gamma^{\mu},\gamma^{5}, \gamma^{5}\gamma^{\mu}, $ and
$\Sigma^{\mu \nu}=i[\gamma^{\mu}, \gamma^{\nu}]$, together with
the metric and the volume 4-form  $\epsilon^{\mu \nu \rho
\sigma}$.  As already mentioned, we are not interested  in probing
the terms involving the Ricci tensor or the scalar curvature, as
they would be expressible using Einstein's equations in terms of
the local energy momentum tensor associated with the matter
fields. In this event the sole term that would be left is
\be
{\cal L}^\prime_{\rm f}=\frac{\xi }{M_{\rm Pl}} \,W_{\mu \nu
\rho \sigma}\,\epsilon^{\mu \nu \rho \sigma}\,\bar \Psi \gamma^{5}
\Psi\, .
\ee
Where  $W_{\mu \nu \rho \sigma}$ stands for the Weyl
tensor.

 However, as it is well known, the Riemann tensor, and thus
also the Weyl tensor has a vanishing totally antisymmetric part,
so $W_{\mu \nu \rho \sigma}\,\epsilon^{\mu \nu \rho \sigma} =0$.
We therefore conclude that there are no terms of this sort where
the suppression factor is just  $1/M_{\rm Pl}$. Then we move to
consider the terms suppressed by more  powers of $M_{\rm Pl}$
these are: In scalar sector we have a term,
\be
{\cal L}^\prime_{\phi}=\frac{\xi }{M_{\rm Pl}^4}\, W^{\mu \nu
\rho \sigma}\, {\rm Tr}\, \Phi\,
\partial_{\mu}\partial_{\nu}\partial_{\rho}\partial_{\sigma}
\Phi
\ee
In the fermion  sector we have a term,
\be
{\cal L}^\prime_{\psi}=\frac{\xi }{M_{\rm Pl}^3}\, W^{\mu \nu
\rho \sigma}\, \bar\Psi
\gamma_{\mu}\gamma_{\nu}\,\partial_{\rho}\partial_{\sigma} \Psi\,
.
\ee
Here in principle, we should have used covariant rather than
partial derivatives, but we use this point to emphasize that we
can choose at each point of space-time locally normal coordinates
constructed so that the Christofell symbols vanish at that
point. In a very small region of space-time which our probe could
be consider to occupy, we would thus have essentially Minkowski
coordinates, and  would in normal circumstances declare that
gravity has been turned off by using appropriately the free
falling recipe to construct local inertial frames. In our scheme
this turning off of gravitation would not  be complete and there
would be a remaining effect due to the fact that space-time was in some
sense granular rather than
continuous and smooth, and that in its bending it (due in the case of
interest the effects of distant matter), the granular structure  was
forced to emerge. This is precisely what we seek to describe in this paper.

In the vector boson  sector, taking into account the requirement
of gauge invariance,  we have a term of the form
\be {\cal L}^\prime_{\rm m}=\frac{\xi }{M_{\rm Pl}^2}\, W^{\mu \nu
\rho \sigma}\, {\rm Tr}\,( F_{\mu \nu} F_{\rho\sigma})
\label{boson} \ee
The pattern seems very clear, the suppression decreases with
increasing  spin of the field. The minimum suppression  given by
the matter content of the standard model of particle physics is
therefore associated with the vector boson sector term
(\ref{boson}) and amongst these the most natural phenomenological
arena would seem to be the one associated with the Maxwell field.

We finalize this section by pointing out that the introduction of
these new lagrangian terms would lead to a change  not only in the
dynamics of matter fields in the presence of gravitation, but also
to a change of the response of  the metric to the presence of
matter fields. However it is easy to see that the latter would be
even more suppressed that the former. To see this let us write the
total action as, \be
  S = \int M_{\rm Pl}^2\; R(g) + L_{{\rm Matt}}(\Psi) +
  \frac{1}{M_{\rm Pl}^n} F(\Psi,
  W)\ ,
\ee where $\Psi$ stands generically for the matter fields of the
standard Model and $W$ stands for the Weyl tensor. The equations
of motion for the matter fields will receive a contribution
proportional to $M_P^{-n}$, while the Einstein tensor would
receive extra contributions proportional to $M_P^{-n-2}$, and thus
we will ignore the latter in the remainder of the manuscript.

\section{An Alternative Approach}
\label{sec:3}

\noindent From the previous analysis we seem to conclude that
there can be no effect of the sort we are interested in, and such
that it is suppressed by the minimal amount that could be
naturally associated with a quantum gravity effect. This is in
sharp contrast with the situation prevailing in the standard
Quantum Gravity Phenomenology  where the Violation of Lorentz
Symmetry could, at least at first sight, be  associated with
dimension five operators, naturally suppressed by one power of the
Planck mass, and where radiative corrections would actually tend
to transfer the violation  to unsuppressed or even enhanced lower
dimensional operators.  Here we want to explore the  possibility
of effects of the  type we have been discussing, with lower
suppression, arising from alternative descriptions of the
spacetime curvature.

We are interested, for phenomenological reasons, in schemes in
which  locally the gravitational degrees of freedom select
preferential directions, planes or other subspaces of spacetime.
This would present us with a scheme in which, in contrast with the
old Quantum Gravity Phenomenology, there is no need to choose an
add hoc direction in spacetime, but where the gravitational
environment would be making the selection of the locally preferred
directions which  could be naively considered as local violations
of absolute Lorentz invariance. Clearly a situation much closer to
the Machian spirit and that of G.R. than that of the old QGP. The
phenomenological motivation is that the situations involving the
breaking of symmetries afford in general a much more promising
experimental scenario, than those that do not, specially dealing
with what could be at most, extremely small effects.
 Needless is to say that this approach is less
natural than the one we investigated  in section~\ref{sec:2}, and
the only remark that we can make in this regard is to note that
the degree of naturalness can be expected to depend on the
language employed, and  that is conceivable that if one were to
use the language most appropriate to describe the quantum
gravitational degrees of freedom -- an issue that is still unknown
as far as we are concerned -- objects, that look unnatural in the
tensorial language appropriate to sub-Planckian phenomena, might
be rather natural, in that, still to de determined, quantum
gravity language. As an illustration of this point, we recall that
in Loop quantum gravity the fundamental degrees of freedom are
associated with holonomies and fluxes, and thus are at some level
fundamentally nonlocal, in contrast with the local degrees of
freedom usually represented in terms of tensor fields.

The idea is then to search for tensors of lower type containing
information about the spacetime curvature. Clearly these have to
contain information already available in the Riemann tensor, but
perhaps not all such information.  We are thus driven to look at
the Weyl tensor, the principal null vectors of the Weyl tensor and
so forth. One approach that seems to yield the kind of terms we
are interested is to consider the Weyl tensor viewed as a tensor
of type $(2,2)$ as a  mapping from the space of antisymmetric
tensors of type $ (0,2)$, $\cal S$  into itself. As is well known
the spacetime metric  endows the six dimensional vector space
$\cal S$ with a pseudo-Riemannian metric of signature $(+++---)$
\cite{wald}. Then the Weyl tensor is a symmetric operator on this
space ${\cal S}$ , which can therefore be diagonalized, and thus
has a complete set of eigenvectors (which are however not
necessarily orthogonal). Let us consider only the eigenvectors
$\Xi^{(i)}$ corresponding to non-vanishing eigenvalues
$\lambda^{(i)}$, by fixing the normalization of these eigenvectors
to be $\pm 1$ (also drop the null eigenvectors)\footnote{In
spacetimes of  astrophysical interest such as the Kerr metric,
this decomposition of the Weyl tensor is indeed possible and can
be explicitly done.}. We will also assume for simplicity, and in
order to avoid possible ambiguities, that all eigenvalues are
different. We can now use the antisymmetric tensors
$\Xi^{(i)}_{\mu \nu}$ and their associated eigenvalues
$\lambda^{(i)}$ to construct the types of Lagrangian terms we are
interested in. In the same spirit as before, searching for terms
linear in these objects, and recalling that the eigenvalues
$\lambda^{(i)} $ have the dimension of the Riemann tensor, we have
the least possible suppressions in each sector as follows:  In
scalar  sector we have a term,
\be
 {\cal L}_{\phi}=\frac{\xi }{M_{\rm Pl}^2}\,  \sum_{i}\, \lambda^{i}
\,\Xi^{(i)}_{\mu \nu} \,{\rm Tr}\, \Phi\,
\partial^{\mu}\partial^{\nu}\, \Phi\, .
\ee
In the fermion  sector we have a term,
\be {\cal L}_{\psi}=\frac{\xi }{M_{\rm Pl}} \, \sum_{i}\,
\lambda^{i} \,\Xi^{(i)}_{\mu \nu} \,\bar\Psi
\gamma^{\mu}\gamma^{\nu}\,\Psi\, .\label{neta}
 \ee
In vector boson sector, taking into account the requirements of
gauge invariance,  we are lead to a term
\be {\cal L}_{\rm m}=\frac{\xi }{M_{\rm Pl}^2}\,   \sum_{i}\,
\lambda^{i}\, \Xi^{(i)}_{\mu \nu} \,{\rm Tr}\,( F^{\mu}_{\rho} F^
{\rho\nu} )\, .
 \ee
Thus the fermions provide the most promising probes, which seems a
fortunate situation, in this scheme.

  As it was pointed out to us, by a charitable soul,  one could also consider coupling
  directly a scalar made out of the standard model fields to an
  appropriate power of a
  scalar constructed out of the Weyl tensor such as
  $(W_{\mu\nu\rho\sigma}W^{\mu\nu\rho\sigma})^{1/2}$.  This
  proposal, while very reasonable, departs slightly from the spirit of our
  attempts to indicate that the space-time structure would naturally
  and locally select preferential directions, in a way that could
  be probed experimentally. Moreover, the lack of any such
  seemingly
  Lorentz invariance violating interactions would tend to make the
  effects
  much harder to detect experimentally. On the other hand this
  line opens the way to consider, despite their seeming
  unnaturalness,
   effects that would not be
  suppressed by any power of the Planck mass, such as
  \be {\cal L}_{\psi}=
  (W_{\mu\nu\rho\sigma}W^{\mu\nu\rho\sigma})^{1/4}\bar\Psi\Psi\,
   . \label{neta2}
 \ee
We will not focuss on this last proposal in this manuscript.

\section{Some motivations from Quantum Gravity}
\label{sec:4}

\noindent In this section we shall give give some heuristic
arguments to the effect that the basic variables and the strategy
followed to get to an effective QFT (on a background) coming from
loop quantum gravity \cite{LoopQG},  and from String Theory, might
allow for the type of terms that we have proposed before.

Let us  first  look at the issue from the LQG perspective. We
start with the observation  that LQG is based on a connection
formulation of General Relativity, where one of the basic objects
is an $SO(3,1)$ connection $A_{\mu C}^B$, whose curvature
$F^{{}\;C}_{\mu\nu B}$ is, on shell, equal to the Riemann tensor.
The second observation is that there is a very close relationship
between the algebraic structure of the Riemann tensor (or the Weyl
tensor in vacuum) and the so called holonomy group of the
spacetime. Basically the idea is that on a given point on
spacetime, the holonomy along small closed curves, yielding
elements of the gauge group (Lorentz) will belong to a specific
subgroup of the Lorentz group \cite{hall}, depending on the
algebraic structure of the curvature tensor. Furthermore, the
proper bivectors $\Xi^{(i)}$ appearing in the expansion
(\ref{neta})  will have information precisely about the holonomies
of the curvature tensor at that point. This is particularly
relevant given that in the LQG formalism, both in the Hamiltonian
approach (where the curvature is now only referring to the spacial
part of the Lorentz group, namely local rotations), and in the
still incomplete Lagrangian formulation (spin foams), the
curvature tensor is always approximated by one of the fundamental
objects in the theory, namely, the (quantum) holonomy.

One can imagine that a detailed treatment of semiclassical states
for the geometry and the regularization of the matter fields on
top of the quantum geometry (assuming that the subtleties and
ambiguities already encountered in the discretization procedure,
are resolved), might yield terms of the form (\ref{neta}). Let us
now briefly discuss how this terms might arise. In the standard
Hamiltonian for a spinor field one has a term of the form, \be
H=\int_\Sigma\d^3\! x\, N E^a_i \frac{1}{\sqrt{{\rm det}\,
q}}\,\left(i\pi^T\tau_i {\cal D}_a\varphi+{\cal D}_a(\pi^T\tau_i
\varphi )+\cdots\right)\, . \ee where $N$ is the lapse function,
${\rm det}\, q$ is the determinant of the spatial metric, $E^a_i$
is the triad field, $ {\cal D}_a = \partial_a + [A_a,\cdot \;]$ is
the covariant derivative, associated with the spatial connection
$A_a$, $\varphi $ is the Weyl spinor corresponding to the fermion
field, $ \pi $ its canonical conjugate, and $\tau_i$ are the Pauli
matrices. A standard trick is to  bring this integral to the form,
\be H=\int\d^3\! x N\left(4\epsilon^{ijk}\epsilon^{abc}\,\frac{\{
A_a^i(x),V(x,\delta)\}\ ,\{
A^j_b,V(x,\delta)\}}{\sqrt{q(x)}}\right)[\pi^T\tau_k{\cal
D}_c\varphi -c.c.]\, , \ee where $V(x,\delta)$ is the volume
element in a finite region. When the expression is turned into an
operator, one replaces the Poisson brackets for commutators and
the connection $A^i_a$ is approximated by holonomies. Finally, the
strategy is to consider semi-classical states $|W,\xi\rangle$ of
the geometry and the matter, and to compute the expectation value
of the Hamiltonian operator $\hat{H}$ on this state to get an
effective Hamiltonian for the spinor field. In the existing works
 on this subject (see \cite{LoopQG}) some simple
assumptions for  the expectation values of certain operators in
the semi-classical state are made, in particular in connection
with the properties of the resulting geometry such as rotational
symmetry of the macroscopic spacetime that corresponds to the
state of the quantum geometry. Our observation is that, in
describing at the fundamental level a region of spacetime which
corresponds macroscopically to a given geometry, one would require
 a sufficiently detailed (and sophisticated)
semiclassical state $|W,\xi\rangle$ that will  contain, perhaps in
a rather convoluted way, information about the extended space-time
structure, and thus it should yield, subleading terms  of the
desired form (\ref{neta}), in the expectation values of complex
operators.

It is, we believe, certainly worth pursuing this avenue, where the
manifestation of the underlying discreteness would not be present
as universal modifications of dispersion relations (closely
related to LIV), but in the appearance of these new terms in the
effective field theory, which would produce them locally. Needless
to say, some work is needed in this direction to make these ideas
precise.

From the String Theory outlook the situation is not clear because
at this time there is no Nonperturbative formulation of the theory
which includes a complete and clear understanding of the emergence
of physical spacetime. However it seems safe to argue that any
such scheme would at some level describe a mildly curved spacetime
in terms of a sufficiently complex state involving many gravitons.
As the spacetime curvature of the situation we envision is
macroscopic, the gravitons would be in some sort of coherent
state. It is well known that the Weyl tensor is intimately
connected with the polarization of a gravitational wave and thus
it will be connected with the common polarization of the gravitons
in the coherent state. In string theory the standard interaction
of gravitons and other matter fields arises at the lowest order in
perturbation theory, but other forms of interaction involving
heavier modes are known to occur, however suppressed by the string
length $\sqrt{\alpha'}$, in the perturbation expansion.  These
higher order interactions will couple the gravitons to matter and
it is quite natural to expect that some of these interactions will
be polarization dependent.  Thus a mechanism to produce an
effective coupling of the spacetime curvature (associated with the
polarization of the gravitons) and suppressed by the string mass
scale which we take to be of the order of $M_{\rm Pl}$ seems to be
in place.

Let us now consider the possible  approaches  for experimental
tests for these type of theories.

\section{Phenomenology}
\label{sec:5}

\noindent The views presented here clearly call for a reassessment
of the phenomenology of quantum gravity. To start with, we are
called to concentrate clearly in the fermion sector as it is the
one leading to the most promisingly observable effects.

Let us write again the corresponding Lagrangian term now taking
into account possible flavor dependence;
\be {\cal L}^{(2)}_{f}=\sum_{a}\frac{\xi_a}{M_{\rm Pl}}\,
\sum_{i}\, \lambda^{i}
 \, \Xi^{(i)}_{\mu \nu}
 \, \bar\Psi_a  \gamma^{\mu}\gamma^{\nu}\Psi_a\, .\label{10}
\ee
where $a$ denotes flavor.  Next we note that we have in principle
the same types of effects that have  been considered in  the
Standard Model Extension (SME) \cite{SME} but only with terms of
the form $-1/2 H_{\mu \nu}\, \bar\Psi \sigma ^{\mu \nu} \Psi$.
Moreover, here  the tensor $H_{\mu \nu}$ must be identified with $
-\frac{2\xi }{M_{\rm Pl}} \sum_{i} \lambda^{i}\, \Xi^{(i)}_{\mu
\nu}$ has a predetermined space-time dependence dictated by the
surrounding gravitational environment.

Thus different experiments at different sites could be compared
only after taking into account the differences in the surrounding
environment that leads to variable values of the relevant
curvature related tensors.

The relevant experiments would thus be associated with both,
relative large gravitational tidal effects (indicating large
curvature)  in the local environment together with either very
high sensitivities in the probes. Furthermore it  is clear that
the probes would need to  involve polarized matter as the explicit
appearance of the Dirac matrix $[\gamma^\mu, \gamma^{\nu}]$
indicates.  Naturally  neutrinos crossing regions of large
curvature such as supernova interiors come to mind. It is
noteworthy that a term of the sort we are considering could lead
to neutrino oscillations even if they are massless, in close
analogy with the ideas exposed in \cite{neutrinos}.

Next we note that  the term  in question does not violate CPT so
that that particular phenomenological avenue is closed. On the
other hand other discrete symmetries, particularly CP could,
depending on the environment and state of motion of the probes be
open channels for investigation.

In the case of potentially flavor dependent effects,  which would
look as violations of the equivalence principle, we must recall
that in the present scheme these are expected only in regions with
large gravitational gradients. Thus the fifth force type tests
\cite{FifthForce}, such as the torsion balance experiments must be
revised. The relevant non-relativistic hamiltonian for a particle
with flavor $a$  can be directly read of from equation (\ref{10})
using the formulation of  \cite{NRH} as
\be
 {\cal H}_{\rm NR}= \epsilon _{jkl} \left[(\half H_{kl}(
\sigma_j + (\vec{\sigma}\cdot \vec{p})\, p_j/m^2) -(1+ \half\,
p^2/m^2) H_{l0}\, \sigma_k \,p_j/m \right]
\ee
  We note that the term proportional to $H_{l0}$
can only come from a non-stationary aspect of the source which in
the case of ground based  experimental setups  case would  seem to
entail the Earth's rotation, and would thus be further reduced in
magnitude. The remaining part would seem to be the most promising
one, and in this case we should clearly  focuss on the first term.

To do this we write this term as
\be
 {\cal H}_{\rm NR}=\xi (1/M_P)  \vec C \cdot
\vec \sigma \ee
where $ C_i = (1/2) \epsilon _{ijk}\sum_{l} \lambda^{l}\,
\Xi^{(l)}_{jk}$ is a sort of magnetic field, which is associated
with the Weyl tensor. As such, in ordinary conditions (static
matter sources) we can expect it to be of order $GM/r^3$, a
feature that indicates that in contrast with the case of  the
ordinary Newtonian gravity, the effects of small and close-by
sources will be of the same relative importance as those of large
but relative distant sources.  In the case of a spherical source,
for instance the effect at its surface, will be determined solely
by its density and not by its size $ |C|\approx (4\pi/3) G\rho$.
This is a serious impediment for constructing  sources that can
lead to large effects.  From this perspective, the most attractive
location  for  a test would be,  the vicinity of a nucleus where
tests such as those designed to look at a possible scale
dependence of the gravitational consatant\cite{AFrank} could be of
use, and also the surface of neutron stars.

In the realm of more ordinary and  Earth-based alternatives, one
needs to note the need to use spin  polarized matter in the probe,
which normally makes the experiments difficult due to magnetic
effects.

To estimate order of magnitude of the effects one would be seeking
in an the experiment we compute the ratio of the contribution to
the energy of the term in question to the ordinary potential
gravitational energy of  the probe.
 That is,
\be \frac{\delta E}{ E} = \frac{\langle \sigma \rangle (\xi/M_{\rm
Pl}) (GM /r^3)}{( GMm/r)} =\xi (l_p l_{\rm Compton}/r^2)  P = \xi
P 10^{-25} \frac{(l_{\rm Compton}/{\rm cm} )}{(r/10^{-4}{\rm
cm})^2}
 \ee
 Where, $r$ is the effective size of the source and $P$ stands for
the mean polarization of matter, $l_{\rm Compton}$ is the Compton
length of the constitutive fermions.\footnote{ We would need to
model the contribution of quarks and electrons to the overall spin
polarization of the probe, and weight their Compton lengths
appropriately} Where we have taken as indicative of what one must
look for,  the value of $r$ to be of  the order of the distances
probed in recent Grenoble neutron experiments \cite{h} $r \approx
10^{-4}$cm. As for the Compton wavelength  we should point out
that, for instance, the neutral Kaon system has a Compton length
associated to the mass difference of the order of 30 cm \cite{l}.
One should then look for systems (like a solid) whose excitations
have an effective mass and thus a Compton length that is large
enough to have a sizable possibility of having observable effects.

Special attention should  be directed into experiments with single
polarized particles (so that $P =1$), in the quantum regime, where
one would look for unexpected phase changes associated with the
polarization and with the presence of gradients of the Newtonian
gravitational acceleration. These again can be expected to be
extremely small, however we should point out that in the limited
experimental evidence there is a report of a rather large anomaly
(one part in $10^3$ violation of the equivalence principle) seen
in interference experiments  probing the gravitational field with
neutrons \cite{NeutrInt} while similar experiments using atoms do
not observe anything despite a higher precision \cite{AtomInt}. In
this case the effect seems to be many orders of magnitude larger
that what we could  expect, and of course, in all such cases of
reported anomalies these must be taken with great care, and
therefore at this point and given its magnitude we are not taking
it as indicative of evidence in  favor of our proposal but rather
just as an example of the kind of effects one might hope to
investigate within this setting.

We end up emphasizing that although, one could think that unless
$\xi $ is unexpectedly large, the outlook  for detecting such
effects is not particularly promising, however we should keep in
mind the lack of relevant experimental results. In particular, one
of the challenges that programs such as Loop Quantum Gravity
should face in the near future is to construct suitable
semiclassical states corresponding to  situations in which large
regions with specific but small curvatures are present. In that
light, any experimental bounds found by a program like the one
suggested here, would play a significant role in guiding the
theoretical developments,  much as the E\"otvos and other early
test of the equivalence principle, played in the construction of
General Relativity.

\section{ Conclusions}
\label{sec:6}

\noindent We have considered a new phenomenological scenario to
search for signatures of quantum gravity.   This new scheme starts
by rejecting from the outset the notion of preferential frames,
thus keeping intact the conceptual framework leading to special
and general relativity.

We must emphasize at this point that  for instance,  loop quantum
gravity, has nothing in its  framework that necessitates the
breakdown of Lorentz Invariance that is associated with the
existence of a preferential reference frame. All that has been
said in this regard up to  this point are heuristic proposals for
states of the theory \cite{LoopQG}, that would result in the types
of effects discussed in \cite{Quant Fluct}. Moreover, given any of
such states, there is nothing in principle preventing the
construction of new states by applying a kind of {\it group
averaging procedure} by means of Lorentz boost (using an
appropriate Loop quantum gravity operator) to the original state.
In this way one could conceive a suitable `superposition' of
states which will not be associated with any preferential frame.
Such type of scenario would be immune to the constraints being set
by the previous explorations of Quantum Gravity Phenomenology.
Furthermore, one might imagine that (in the case one could
construct) a quantum state that satisfies the full set of
constraints, being thus fully diffeomorphism invariant, would be
in an appropriate sense necessarily devoid of the problem of
emergence of preferential frames.

The proposal here is clearly beyond anything that has been
considered so far,  and the search for  experiments that could
provide relevant information seem from the onset to be very
challenging indeed. We hope however  that the ideas proposed here
will motivate our theoretical and experimental colleagues to
explore these intriguing possibilities, either trough some of the
directions we have mentioned here, or trough some new and
cleverly devised experimental set-ups.

\section*{Acknowledgments}

\noindent We thank A. G\"uijosa for helpful discussions. D.S.
acknowledges inspiration from old discussions with E. Fishbach. We
would like to thank the referee and the editor for useful comments
and references. This work was in part supported by DGAPA-UNAM
IN108103 and CONACyT 43914-F grants.

\end{document}